\documentclass[preprint]{revtex4}
\usepackage{amssymb,amsmath}

\begin{document}

\draft
\title{ Quantitative Resolution to some "Absolute Discrepancies" in Cancer Theories \\
  a View from Phage lambda Genetic Switch }
\author{ Ping Ao }
\address{ Department of Mechanical Engineering and Department of Physics \\
               University of Washington, Seattle, WA 98195, USA  }
\date{April 3 (2007) }

\begin{abstract}

Is it possible to understand cancer? Or more specifically, is it
possible to understand cancer from genetic side? There already
many answers in literature. The most optimistic one has claimed
that it is mission-possible. Duesberg and his colleagues reviewed
the impressive amount of research results on cancer accumulated
over 100 years. It confirms the a general opinion that considering
all available experimental results and clinical observations there
is no cancer theory without major difficulties, including the
prevailing gene-based cancer theories. They have then listed 9
"absolute discrepancies" for such cancer theory. In this letter
the quantitative evidence against one of their major reasons for
dismissing mutation cancer theory, by both in vivo
experiment and a first principle computation, is explicitly pointed out. \\
{To cite: P. Ao, 2007, {Orders of magnitude change in phenotype
rate caused by mutation}, \\ Cellular Oncology {\bf  29}: 67 - 69.
}
\end{abstract}

\maketitle



In a forceful article Duesberg and his colleagues (2005) reviewed
the impressive amount of research results on cancer accumulated
over 100 years. It confirms the current conclusion that
considering all available experimental results and clinical
observations there is no cancer theory without major difficulties,
including the prevailing gene-based cancer theories (Hanahan and
Weinberg 2000; Prehn 2002; Vogelstein and Kinzler 2004; Beckman
and Loeb 2005). Phrasing differently, any known cancer theory is
refutable to a substantial degree. While the support for their own
advocated chromosomal cancer theory appears strong, it seems that
their wholesale criticism on mutation cancer theory is premature.
In this letter the quantitative evidence against one of their
major reasons for dismissing mutation cancer theory, by both in
vivo experiment and a first principle computation, is explicitly
pointed out.

Duesberg et al (2005) listed 9 "absolute discrepancies" or
questions which they believe the mutation cancer theory cannot
answer: \\
"(1) How would non-mutagenic carcinogens cause cancer? \\
(2) What kind of mutation would cause cancer only after delays of
several decades and many cell generations? \\
(3) What kind of mutation would alter the phenotype of mutant
cells perpetually, despite the absence of further mutagens? \\
(4) What kind of mutation would be able to alter phenotypes at
rates that exceed conventional gene mutations 4-11 orders of
magnitude? \\
(5) What kind of mutation would generate resistance against many
more drugs than the one used to select it? \\
(6) What kind of mutations would change the cellular and nuclear
morphologies several-fold within the same "clonal" cancer? \\
(7) What kind of mutation would alter the expressions and
metabolic activities of 1000s of genes, which is the hallmark of
cancer cells? \\
(8) What kind of mutation would consistently coincide with
aneuploidy, although conventional gene mutations generate infinite
numbers of new phenotypes without altering the karyotype? \\
(9) Why would cancer not be heritable via conventional mutations
by conventional Mendelian genetics?"

Evidently those important questions should be considered seriously
by any cancer researcher. There are, however, two general reasons
that the temporally inability to address them is not enough to
dismiss mutation cancer theory. First, in many situations whether
the effects are the causes or the consequences, or mutual causes
to each other, are still poorly understood, though their
associations with cancers may be obvious. The clarification of
such confusing requires further and more experimental and clinical
studies. Such efforts have been carrying out, for example, by
Weiss et al (2004), Hermsen et al (2005), Weber et al (2006);
Bielas et al (2006), Levitus et al (2006), and Sjoblom et al
(2006).

Second, the absence of an explanation or of a theory is not a
proof that it would not ever exist. If there were already enough
amount of consistent experimental and clinical observations, the
emergence of a theory would be simply a matter of time. It is a
test to our creativity and imagination. Therefore, those "absolute
discrepancies" are logically not necessarily against the mutation
cancer theory. Having given a "dodged" defense, here I would like
to call the attention to one evidence specifically addressing
above 4th discrepancy: "(4) What kind of mutation would be able to
alter phenotypes at rates that exceed conventional gene mutations
4-11 orders of magnitude?" The question (4) can be answered by
mutation cancer theory in a quantitative and first principle
manner. The quantitative evidence also suggests answers to
questions (1) and (2) of Duesberg et al.

In the phage lambda system, a bio-system arguably started the
modern molecular biology (Cairns et al 1992), extensive and
quantitative experiments have demonstrated that simple mutations
can cause the change in phenotypes over many orders of magnitude
(Ptashne 2004; Oppenheim et al 2005). The phenotype easily
accessible to experimental study is the switching between
lysogenic and lytic states. It is generally known that a mutation
in the DNA binding sites can cause more than 100 and more folds
change in the switching rate, controlled by a few base pairs in
the genomic sequences (Revet et al 1999; Little et al 1999; Dodd
et al 2005). Systematic study showed that at least over 8 orders
of magnitude change can be observed among mutants.  A quantitative
study is summarized in Table I.

{\ }

\noindent
\begin{tabular}{|c|c|c|c|c|c|}
  \hline
  Phage lambda genotype & $\lambda^{+}$  ($\lambda^{+}
  O_R$321, wild) & $\lambda^{+} O_R$3'23' & $\lambda^{+} O_R$121
  & $\lambda^{+} O_R$323 & $\lambda^{+} O_R$123 \\
  \hline
  Switching rate (exper) & $2\times 10^{-9}$ &  $5\times 10^{-7}$
  &  $3\times 10^{-6}$ &  $2\times 10^{-5}$ & $\infty$ \\
  Switching rate (theor) & $1\times 10^{-9}$ & $1\times
  10^{-7}$

  & $3\times 10^{-6}$ & $7\times 10^{-5}$ & $\infty$ \\
  \hline
\end{tabular}

{\ }

Table I:  This table is based on Zhu et al (2004, 2006). There are
5 different phage lambda phenotypes, including the "wild type",
which have been systematic studied experimentally (Little et al
1999). The switching rate is the probability to switch from
lysogenic to lytic states per minute under normal lab condition.
The symbol ? indicates that there is no stable lysogenic state,
that is, the phage lambda would immediately switch to lytic state.
The switching rate is then "infinite".

{\ }

\noindent
Quantitative answer to "absolute discrepancy" 4: \\
The point should be emphasized is that the mathematical
calculation is based on first principle modelling without "free"
parameters. What the "first principle" means is that the
interaction between involved proteins and the protein-DNA binding
are based on carefully reasoned physical, chemical and biological
principles during past 40 years. What the "parameter free" means
is that all the kinetic parameters needed for the mathematical
modelling have been fixed by other experiments. Thus, the
remarkable consistency over at least 8 orders of magnitude between
the experimental data and mathematical calculation shows that it
is unlikely due to artifacts in experiment and/or in modelling.
Because such effect can occur in phage lambda, there is no reason
that same thing cannot occur in higher organisms (Ptashne and
Gann, 2002): Numerous gene regulatory sites similar to that of
phage lambda exist in our human genome and wrong switching in gene
regulatory network is generally believed to contribute to cancer.
Therefore, the answer to the question (4) of Duesberg et al based
on mutation cancer theory is already positive.

{\ }

\noindent
Quantitative answer to "absolute discrepancy" 2: \\
The viability of various mutants, some can live up to thousands of
generations before going to lytic state to kill its host E. coli,
suggests that there can be a long delay in the manifestation in
phenotypes after a mutation.  Such a gene regulatory example hence
directly answers the question (2) of Duesberg et al.

{\ }

\noindent
Quantitative answer to "absolute discrepancy" 1: \\
In addition, it is known that the stability of lambda genetic
switch can be influenced both chemically and physically, without
any mutagenic effect (Ptashne 2004; Zhu et al 2004), that is,
non-mutagenic agents can cause the switch  from lysogenic to lytic
states, therefore changes the an otherwise robust phenotype. This
fact suggests itself as an answer to the question (1) of Duesberg
et al.

To summarize, though whether chromosomal or mutation cancer
theory, or both , are the candidates for the cancer theory is too
early to call, Duesberg et al is premature to write out the
mutation cancer theory. Even if neither were the final cancer
theory, both already appear clinically relevant (Meijer, 2005) and
should be studied thoroughly. Finally, I would like to venture a
challenge to my fellow quantitative modelers: Is it possible to
address all those "absolute discrepancy" quantitatively? I believe
you can do better than what presented here.

{\ }

I thank R. Prehn and L. Loeb for stimulating discussions. This
work was supported in part by US National Institutes of Health
under grant number HG002894.

{\ }


\end{document}